\documentclass[aps,twocolumn,floats,prd,nofootinbib,superscriptaddress,10pt]{revtex4-2}

\def\HI{{\rm H\,{\footnotesize I}~}}

\def\Ts{T_{\rm S}}
\def\Tk{T_{\rm K}}
\def\Tcmb{T_{\rm CMB}}

\def\lya{Lyman-$\alpha$ }
\def\mpci{~{\rm Mpc}^{-1}}

\def\be{\begin{equation}}
\def\ee{\end{equation}}

\def\kpa{k_{\parallel}}
\def\kpe{k_{\perp}}

\def\ca{{Scenario A}}
\def\cb{{Scenario B}}
\def\cc{{Scenario C}}
\def\cai{{Scenario A-I}}
\def\caii{{Scenario A-II}}
\def\caiii{{Scenario A-III}}
\def\21cmf{\texttt{21cmFAST}}

\makeatletter
\def\hlinewd#1{%
\noalign{\ifnum0=`}\fi\hrule \@height #1 \futurelet
\reserved@a\@xhline}
\makeatother

\usepackage{pifont}

\usepackage{amsmath}
\usepackage{amssymb}
\usepackage{graphicx}
\usepackage{multirow}
\usepackage{makecell}
\usepackage{hyperref}
\usepackage{adjustbox}
\usepackage{tabularx}
\hypersetup{
	colorlinks=true,
	linkcolor=red,
	filecolor=magenta,      
	urlcolor=blue,
}
\usepackage[normalem]{ulem}
\usepackage[usenames,dvipsnames]{xcolor} 
\usepackage{comment}
\usepackage{makecell}
\usepackage{subfigure}
\usepackage{placeins}

\begin{document}

\title{Mitigating the optical depth degeneracy in the cosmological measurement of neutrino masses using 21-cm observations}

\author{Gali Shmueli}
\email{shmugal@post.bgu.ac.il}
\affiliation{Department of Physics, Ben-Gurion University Be'er Sheva 84105, Israel}
\author{Debanjan Sarkar}
\email{debanjan@post.bgu.ac.il}
\affiliation{Department of Physics, Ben-Gurion University Be'er Sheva 84105, Israel}
\author{Ely D. Kovetz}
\email{kovetz@bgu.ac.il}
\affiliation{Department of Physics, Ben-Gurion University Be'er Sheva 84105, Israel}

\date{\today}

\begin{abstract}
Massive neutrinos modify the expansion history of the universe and suppress the structure formation below their 
free streaming scale. Cosmic microwave background (CMB) observations 
at small angular scales can be used to constrain the total mass $\Sigma m_\nu$ of the three neutrino flavors. 
However, at these scales, the CMB-measured $\Sigma m_\nu$ is degenerate with $\tau$, the optical depth to reionization,
which quantifies the damping of CMB anisotropies due to the scattering of CMB photons with free electrons along the line of sight. 
Here we revisit the idea to use 21-cm power spectrum observations to provide direct estimates for $\tau$. 
A joint analysis of CMB and 21-cm data can alleviate the $\tau-\Sigma m_\nu$ degeneracy, making it possible to
measure $\Sigma m_\nu$ with unprecedented precision. Forecasting for the upcoming 
Hydrogen Epoch of Reionization Array (HERA), we find that a $\lesssim\mathcal{O}(10\%)$ measurement of $\tau$ is achievable, 
which would enable a $\gtrsim  5\sigma$ measurement of $\Sigma m_\nu=60\,[{\rm meV}]$, for any
astrophysics model that we considered. Precise estimates of $\tau$ also help  reduce uncertainties in  other cosmological parameters, 
such as $A_s$, the amplitude of the primordial scalar fluctuations power spectrum. \end{abstract}

\maketitle

\section{Introduction}\label{sec:intro}

Since its discovery, the cosmic microwave background (CMB) \cite{Gawiser:2000az} has played a dominant role 
in our understanding of the Universe. Observing the CMB allows us to learn more about the origins 
and evolution of the universe, and test our current understanding of fundamental physics
\cite{Feng:2004mq,Challinor:2006yh,Li:2008aia, Hu:2008hd,Li:2008tma,Gubitosi:2009eu,
Challinor:2012ws,Fedderke:2019ajk,Komatsu:2022nvu}. 
One such example is the measurement of the sum of neutrino masses $\sum m_{\nu}$. Neutrinos come in three flavours.
Neutrino oscillation experiments have revealed that neutrinos have mass and obey three possible hierarchies: 
normal, inverted, and degenerate \cite{Super-Kamiokande:1998kpq,Super-Kamiokande:2001bfk,Super-Kamiokande:2002ujc,Soudan2:2003qqa,Drexlin:2003fc,KamLAND:2004mhv,Super-Kamiokande:2005mbp,Messier:2006yg}. 
Due to their non-zero mass, neutrinos contribute to the total energy density of the Universe and affect the cosmic expansion rate and  evolution of cosmic structures~\cite{Eisenstein:1997jh, Lesgourgues:2006nd, Brandbyge:2010ge, Marulli:2011he}.
This renders the CMB and  large-scale structure sensitive probes of the sum of neutrino masses
\cite{Hu:1997mj,Elgaroy:2004rc,Wang:2005vr,Goobar:2006xz,Gratton:2007tb,Reid:2009nq,Pan:2015bgi,
Allison:2015qca,Abazajian:2016hbv,RoyChoudhury:2019hls,Ivanov:2019hqk,Tanseri:2022zfe,Sakr:2022ans}.

However, processes in the late time universe, like the epoch of reionization (EoR) 
\cite{Zaroubi:2012in,Wise:2019qtq}, limit the precision with which we
can measure neutrino masses. Free electrons along the light of sight to the surface of last scattering influence the CMB anisotropies, an 
 effect characterized by the parameter $\tau$---known as the optical depth to reionization~\cite{Keating:2005ds}---which is one of
the six parameters of the $\Lambda$CDM.  
It has two main effects on the CMB power spectra \cite{Haiman:1999me,Hu:1999vq}.
First, it damps the scalar perturbations as generated at recombination by a factor $\exp(-2\tau)$. This makes it
highly degenerate with $A_s$, the amplitude of the primordial scalar perturbations, and at high multipoles, or smaller scales, 
also highly degenerate with the sum of neutrino masses $\sum m_{\nu}$~\cite{Reichardt:2015cos}. 
Secondly, the re-scattering of the CMB photons off
free electrons at the reionization epoch generates a bump in the CMB polarization power spectra 
at large angular scales \cite{WMAP:2003ggs,WMAP:2012nax}. 
Observation of the large-scale CMB polarization thus provides a measurement of $\tau$, which
allows breaking the degeneracy with $A_s$ and $\sum m_{\nu}$ to some extent. 
Improving the measurement of $\tau$ will be crucial for differentiating between the mass hierarchies of neutrinos and enabling a robust detection of the total mass. 

A number of current probes, including the \lya forest \cite{Weinberg:2003eg,Wyithe:2004jw,Fan:2005es}, 
\lya emitting galaxies \cite{Malhotra:2004ef,Jensen:2012uk,Dijkstra:2014xta,Dijkstra:2015jdy},
the kinematic Sunyaev-Zeldovich effect \cite{Ma:2001xr,McQuinn:2005ce,Park:2013mv,Gorce:2020pcy,Gorce:2022cvb},
etc., allow to place direct constraints on  
$\tau$. Similarly, $\sum m_{\nu}$ can also be determined by measuring
the expansion rate using distance ladders \cite{Wyman:2013lza,Riess:2021jrx}, 
from large-scale structure surveys \cite{Villaescusa-Navarro:2017mfx,Palanque-Delabrouille:2019iyz, Ivanov:2019hqk, Chudaykin:2019ock}, 
\lya forest surveys \cite{Palanque-Delabrouille:2014jca,Baur:2017fxd}, 
line-intensity mapping \cite{MoradinezhadDizgah:2021upg,Bernal:2022jap,Libanore:2022ntl},
the post-reionization 21-cm signal \cite{Villaescusa-Navarro:2015cca,Oyama:2015gma,Pal:2016icc,Sarkar:2016lvb,
Sarkar:2018gcb,Sarkar:2019nak,Sarkar:2019ojl},
etc. All these independent observations can be combined together for a precision measurement of 
$\sum m_{\nu}$. In this paper, we assess the feasibility of using the 21-cm signal from 
EoR as a direct probe of $\tau$.

The redshifted 21-cm observations are a very sensitive probe of EoR. A number of telescopes such as
LOFAR~\cite{vanHaarlem:2013dsa}, HERA~\cite{DeBoer:2016tnn}, GMRT~\cite{Paciga:2013fj} and 
SKA~\cite{Ghara:2016dva} are seeking the signal from high redshifts. Fluctuations in the 21-cm 
signal probe the density-weighted electron fraction in the EoR, which is the primary ingredient required to 
compute $\tau$. Therefore, 21-cm observations provide an independent measurement of $\tau$~\cite{Pritchard:2009nm,Liu:2015txa}.
Small-scale damping in the 21-cm power spectrum, caused by the neutrinos, also provides an independent
estimate of $\sum m_{\nu}$~\cite{Pritchard:2008wy}. 
A joint analysis of the 21-cm fluctuations and the CMB data thus helps to 
precisely estimate $\tau$, break the $\tau-\sum m_{\nu}$ degeneracy, and significantly
reduce $\Delta \sum m_{\nu}$. This idea was first coined by Ref. \cite{Liu:2015txa} where it was shown that 
$\Delta \sum m_{\nu}$
can be measured with $\pm 12\,[{\rm meV}]$ accuracy if 21-cm observations from HERA are combined with the CMB observations. 
Here we revisit the analysis presented in Ref.~\cite{Liu:2015txa} with an updated treatment and provide forecasts for the combination of HERA and the planned CMB-S4 experiment~\cite{CMB-S4:2016ple,Wu:2014hta}. 

In order to generate the mock observations of HERA, Ref.~\cite{Liu:2015txa} used the publicly 
available code \21cmf\footnote{\href{https://github.com/21cmfast/21cmFAST}{https://github.com/21cmfast/21cmFAST}.} 
which assumes an inside-out model of 
reionization based on the excursion set formalism 
as described in Ref.~\cite{Mesinger:2010ne}. However, since their analysis, the \21cmf code has been augmented with improved modeling (involving new parametrization) of the cosmic dawn and reionization astrophysics~\cite{Park:2018ljd,Qin:2020xyh,Munoz:2021psm}.
 Furthermore, for simplicity, 
they assumed throughout the calculation that the spin temperature, $\Ts$, is much higher than the CMB temperature, $\Tcmb$ ($\Ts >> \Tcmb$), which is true only
at the very advanced stages of reionization. In addition, they assumed an observation from HERA
spanning a limited redshift range of $6.1 \le z \le 9.1$.

In the analysis presented below, we use the latest version of the \21cmf code, relax the assumption $\Ts >> \Tcmb$, 
 accounting for the exact evolution of $\Ts$, $\Tcmb$ and the kinetic temperature $\Tk$ as computed
by the \21cmf code, and consider  21-cm observations from HERA across the redshift range $5 \le z \le 27$, which
spans the cosmic-dawn (CD) era down to the end of reionization. 
We have further considered additional effects like the 
\lya heating of the inter-galactic medium\footnote{This mechanism is due to the resonant
scattering between \lya photons and the IGM atoms, and is important when the X-ray heating efficiency is not
very high.}~\cite{Chuzhoy:2006au, Chen:2003gc, Oklopcic:2013nda, Ciardi:2009zd, Mittal:2020kjs}, 
Population III stars\footnote{These first generations of stars are assumed to have formed inside
 {\it mini} halos, in the mass range $10^5-10^6\,{\rm M_{\odot}}$, where the molecular cooling process
makes star formation possible in those halos.}~\cite{OShea:2007ita,Mirocha:2017xxz,Trenti:2010hs,Qin:2020xyh,Munoz:2021psm}, 
the relative velocity between dark matter and baryon fluid\footnote{This supersonic relative velocity 
between dark matter and baryons after recombination is generated due to the interaction between 
baryons and photons before recombination. The same interaction gives rise to the baryon acoustic oscillations.
This supersonic velocity applies negative feedback and hinders structure formation inside the mini-halos, modulating star-formation on large scales.~\cite{Ciardi:2005gc,Kimm:2016kkj,Qin:2020xyh}.}~\cite{Fialkov:2014rba, Barkana:2016nyr, Tseliakhovich:2010bj, Bovy:2012af, Stacy:2010gg, Fialkov:2011iw, Schmidt:2016coo}, 
and in-homogeneous Lyman-Werner (LW) radiation feedback\footnote{The UV photons in the Lyman-Werner band ($11.2-13.6$ eV)
photo-dissociate the molecular hydrogen and imposes negative feedback on star formation. In an earlier work,
Ref.~\cite{Munoz:2019rhi} have implemented the LW radiation feedback for the 21-cm calculations 
in the \21cmf code. This work, however, assumed that the intensity of LW radiation does not vary 
spatially, which is physically less plausible. 
In a recent work, Ref.~\cite{Munoz:2021psm} has updated the calculations to include the 
spatial variation of the LW radiation intensity in the latest version of the \21cmf code. 
}~\cite{Fialkov:2012su, Visbal:2014fta, Safranek-Shrader:2012zig, Ricotti:2000at, Haiman:1996rc, Ahn:2008uwe}. 
Note that, in the current public version of the \21cmf code, the \lya heating is not
included. We have made necessary changes in the code to accommodate the \lya heating in our calculations~\cite{Sarkar:2022dvl}. 
Further, we have interfaced \21cmf with the public Boltzmann code 
\texttt{CLASS}\footnote{\href{https://github.com/lesgourg/class\_public}{https://github.com/lesgourg/class\_public}.}~
\cite{Lesgourgues:2011re,Blas:2011rf,Lesgourgues:2011rg,Lesgourgues:2011rh}
so that the cosmological and astrophysical parameter degeneracies can be studied consistently in a joint analysis of the CMB and 21-cm signals. For more details on our 
implementation, the reader is referred to Refs.~\cite{Sarkar:2022dvl,Sarkar:2022mdz}.

Using Fisher-based forecasts, we demonstrate that the combination of 21-cm and CMB data from HERA and CMB-S4 can yield a measurement of $\sum m_{\nu}$ beyond the precision required for a robust determination of the neutrino mass hierarchy and a greater than $5\sigma$ detection of the minimal sum of neutrino masses.

This paper is structured as follows.
In Section~\ref{sec:exps}, we describe the fiducial experiments and observables used in our analysis, and
present in detail the methodology used in this work. In Sections~\ref{sec:CMBresults} and~\ref{sec:scenarios} we present our results,
and in Section~\ref{sec:summary} we summarize our main findings.

\section{Formalism} 
\label{sec:exps}

 In this section, we discuss the fiducial experiments we chose to consider, along with the various observables 
 as well as the assumptions made throughout the analysis.

\subsection{CMB}
\label{sec:cmb_obs}
For the CMB experiments, we mainly used the publicly available data products from $Planck$-2018 data release~\cite{Planck:2018nkj,Planck:2018vyg}. 
We focus on the ``TT, TE, EE + LowE + Lensing + BAO" dataset since it provides the tightest errors on parameters. 
We use the best-fit values from this dataset as our fiducial  cosmological parameters, 
and use the covariance matrices provided by the $Planck$ collaboration 
\footnote{\href{https://wiki.cosmos.esa.int/planck-legacy-archive/index.php/Cosmological\_Parameters}{https://wiki.cosmos.esa.int/planck-legacy-archive/index.php/Cosmological\_Parameters}}. 
For simplicity, we assume Gaussian parameter uncertainties so that the inverse of the covariance
matrix gives us the Fisher matrix which we require later in the analysis.
More details are given in Appendix~\ref{sec:cmbs4}.

\subsection{21-cm Observations}
\label{sec:21cm_obs}
Next, we consider radio-interferometer observations of the redshifted 21-cm signal from the neutral hydrogen (HI). These  measure the Fourier transform of the two-point correlation function of the intensity fluctuations, also known as the power spectrum
\begin{equation}
    \Delta_{21}^2(k,z) = \frac{k^3 P_{21}(k,z)}{2\pi^2}
\end{equation}
where $P_{21}(k,z) = \langle \Tilde{T}_{21}(\Vec{k},z) \Tilde{T}_{21}^*(\Vec{k},z) \rangle$ and $\Tilde{T}_{21}(\Vec{k},z)$ is the Fourier transform of $T_{21}(\Vec{x},z) - \langle T_{21}(z) \rangle$. Here $T_{21}(\Vec{x},z)$ represents
the 21-cm intensity fluctuations from various positions and directions in the sky, and $\langle T_{21}(z) \rangle$ 
is the sky average of the intensity fluctuations at a single redshift. $T_{21}(\Vec{x},z)$ can be approximately
written as~\cite{Madau:1996cs, Barkana:2000fd, Bharadwaj:2004it},
\begin{equation}
    \label{eq:T21}
    T_{21}(\Vec{x},z) \approx T_0 x_{\rm HI} (1+\delta_b) \left( 1-\frac{T_{\rm CMB}}{T_s} \right) \left(\frac{H}{H+\partial v_r /\partial r} \right) \,,
\end{equation}
with $T_0$ being,
\begin{equation}
    T_0 = \frac{9 \hbar c^2 A_{10} \Omega_b H_0}{128 \pi G k_B \nu^2_{21}\mu m_p \Omega^{1/2}_m} \left( 1 - \frac{Y^{\rm BBN}_p}{4} \right)
\end{equation}
where $\hbar$ is Planck's constant, $A_{10} = 2.86 \times 10^{-15} s^{-1}$ is the Einstein coefficient for the hyperfine (21-cm) transition, $G$ is the gravitational constant, $k_B$ is Boltmann's constant, $\nu_{21} \approx 1420$ MHz is the frequency of the 21-cm line, $m_p$ is the proton mass, $H_0$ is the Hubble parameter, $\Omega_b$ is the normalized baryon energy density, $\Omega_m$ is the normalized matter density, $Y^{\rm BBN}_p$ is the helium abundance and $\mu$ is the mean molecular weight.

In order to simulate the observed 21-cm power spectrum and global signal for an input set of astrophysical and cosmological parameters, we use a modified version of the public 
semi-numerical code \21cmf~\cite{Munoz:2021psm}\footnote{\href{https://github.com/debanjan-cosmo/21cmFAST/tree/21cmFAST-heating}{https://github.com/debanjan-cosmo/21cmFAST/tree/21cmFAST-heating}}, used by Refs.~\cite{Sarkar:2022dvl,Sarkar:2022mdz}, that includes the effects of \lya. This code is also interfaced with the \texttt{CLASS} code so that the degeneracies of the cosmological
parameters can be robustly studied. Once again, for more information about the various features used in our code, the 
reader is referred to Refs.~\cite{Sarkar:2022dvl, Sarkar:2022mdz}. The astrophysical parameters and modeling
we used are elaborated on in Refs.~\cite{Qin:2020xyh,Munoz:2021psm}.

We have considered three different configurations of the \21cmf code in calculating our results:

$\bullet$ \textbf{\ca}: Here we assume that both the Population II stars (formed inside the halos that contain
 atomic-cooling galaxies or \textbf{ACG}s\footnote{ACGs mainly obtained their gas through \HI (and
He) line transitions that are efficient at virial temperatures $T_{\rm vir}\gtrsim10^4\,{\rm K}$.}) and Population III stars (formed inside mini-halos containing
molecular-cooling galaxies \textbf{MCG}s\footnote{Inside MCGs, the gas cools mainly
through the ${\rm H_2}$ rotational–vibrational transitions efficient at
$T_{\rm vir} \sim 10^3 - 10^4\,{\rm K}$. Note that, most ACGs at high redshifts are “second-generation” galaxies, forming out of MCGs.})
are present in the simulations. The presence of Population III prepones the onset of cosmic dawn, 
which otherwise would occur at a later time~
\cite{Qin:2020xyh,Munoz:2021psm}. Population III stars affect the evolution
of the 21-cm signal significantly, as can be seen in various recent works
\cite{Fialkov:2014rba, Barkana:2016nyr,Bovy:2012af, Tseliakhovich:2010bj, Stacy:2010gg,
Fialkov:2011iw, Schmidt:2016coo, Munoz:2019rhi}.
In Table~\ref{table:astro_params_description}, we mention some of the main astrophysical parameters
used in this work.

\begin{table*}[]
\begin{ruledtabular}
\begin{tabularx}{\textwidth}{ll}
{\bf Parameters} & {\bf Description} \\ \hline
$f_{\star,10}$ \dotfill & Stellar to halo mass ratio at $M_{\rm vir}=10^{10}\,{\rm M}_{\odot}$ for ACGs\\ \hline
$f_{\star,7}$ \dotfill & Stellar to halo mass ratio at $M_{\rm vir}=10^{7}\,{\rm M}_{\odot}$ for MCGs \\ \hline
$f_{\rm esc,10}$ \dotfill & Escape fraction of ionizing photons at $M_{\rm vir}=10^{10}\,{\rm M}_{\odot}$ for ACGs\\ \hline
$f_{\rm esc,7}$ \dotfill & Escape fraction of ionizing photons at $M_{\rm vir}=10^{7}\,{\rm M}_{\odot}$ for MCGs\\ \hline
$L_X$ \dotfill & Soft-band X-ray luminosity per SFR in units of ${\rm erg} \;{\rm s}^{-1}\;{\rm M}_{\odot}^{-1}\;{\rm yr}$
for ACGs \\ \hline
$L_{\rm X,mini}$ \dotfill & Soft-band X-ray luminosity per SFR in units of ${\rm erg} \;{\rm s}^{-1}\;{\rm M}_{\odot}^{-1}\;{\rm yr}$ for MCGs \\
\end{tabularx}
\end{ruledtabular}
\caption{The main astrophysics parameters and their definitions used in
\ca~and \cb~(refer to Section~\ref{sec:21cm_obs}).}
\label{table:astro_params_description}
\end{table*}

$\bullet$ \textbf{\cb}: Here we do not consider the contribution from  MCGs, retaining only 
 ACGs in the simulations. 

$\bullet$ \textbf{\cc}: In this case, we adopt the old parametrization of the \21cmf code 
as was done in Ref.~\cite{Liu:2015txa}. Following Ref.~\cite{Liu:2015txa}, we use three parameters to parametrize the reionization process: 
$T_{\rm vir}$, the minimum virial temperature 
of the first ionizing galaxies; $\zeta$, the ionizing efficiency of those galaxies; and 
$R_{\rm mfp}$,  the mean free path of ionizing photons in ionized regions
in the Universe. 
This lets us compare our estimates directly with the findings 
of Ref.~\cite{Liu:2015txa}. This old parametrization, like in \cb, includes only ACGs. 
However, note that we can only compare our results qualitatively with Ref.~\cite{Liu:2015txa} 
as we are using the latest version of the \21cmf code where some calculations 
have been updated.

It is important to note that, \ca~is believed to represent the most updated prescription of astrophysics
at high redshifts, and we shall mainly focus on \ca~when presenting our results.
Meanwhile, \cb~and \cc~will be considered as two alternative simple prescriptions to be used
for comparison. 

For all three scenarios, we run  the \21cmf code with a box
size of $600$ Mpc and $1$ Mpc resolution to compute the 21-cm global signal and fluctuations.
We drop the assumption $\Ts >> \Tcmb$ used in Ref.~\cite{Liu:2015txa} and consider the exact evolution of the 
temperatures as given by \21cmf. We also consider the \lya heating, 
the relative velocity between dark matter and baryon, and 
regular LW radiation feedback strength (as defined in Ref.~\cite{Munoz:2021psm}) for all the simulations. 
Note that, the relative velocity and the LW feedback mostly affect the MCGs. We, therefore, expect to 
see the signatures of these effects only in \ca.

We make forecasts for the HERA 21-cm intensity mapping 
experiment~\cite{DeBoer:2016tnn}. 
HERA will measure the 21-cm fluctuations 
from Cosmic Dawn ($50$ MHz or $z\sim27$) to the reionization era ($225$ MHz or $z\sim5$). 
The ultimate setup of HERA is expected to contain $350$
antenna dishes, each with a diameter of $14$ m. Out of the $350$ dishes, $320$ will be 
placed in a close-packed hexagonal configuration and the remaining $30$ will be placed at longer 
baselines. We calculate the sensitivity of the HERA observations using the publicly available
package \texttt{21cmSense}
\footnote{\href{https://github.com/steven-murray/21cmSense}{github.com/steven-murray/21cmSense}}
~\cite{Pober:2012zz,Pober:2013jna}. 
This code accounts for the $u-v$ sensitivities of each antenna
in the array and calculates the possible errors in the 21-cm power spectrum measurement,
including cosmic variance. 
The total redshift coverage of HERA is divided into 30 bins, and we assume that all the 
redshift bins are observed simultaneously for a total of $180$ days with 6 hours of observation per day.  For the receiver temperature we take $T_{\rm rec} = 100\,{\rm K}$. 

In the 21-cm observations, the foreground is many orders of magnitude brighter than the signal and is 
anticipated to contaminate a significant amount of Fourier space
\cite{Bowman:2008mk,Dillon:2012wx,Hazelton:2013xu,Liu:2011hh}. 
In the $\kpe-\kpa$ space, where $\kpa$ and $\kpe$ are the components
of the wave vector respectively parallel and perpendicular to the line-of-sight direction, the contaminated part of
the Fourier space appears like a ``wedge'' \cite{Datta:2010pk,Pober:2013jna}. The extent of this foreground wedge
can be parametrized \cite{Pober:2012zz,Pober:2013jna} by assuming that all wave numbers with $\kpa$ below
\begin{equation*}
    \kpa^{\rm min} = a + b(z)\kpe\,,
\end{equation*}
are contaminated, where $b(z)$ accounts for the chromaticity of the antennae, and $a$ is a constant superhorizon
buffer. In this paper, we consider ``moderate'' foreground contamination 
(as defined in Refs.~\cite{Pober:2012zz,Pober:2013jna} )
in \texttt{21cmSense} for the 21-cm observations with HERA.
Here we assume $b(z)$ is determined by the horizon limit, and consider $a = 0.05\,h\mpci$.
Further, the baselines are added coherently.
For a comprehensive reading about the details of the different setups and foreground scenarios, 
the reader is referred to Refs.~\cite{Pober:2012zz,Pober:2013jna} .

\subsection{Optical depth calculation}\label{sec:tauCalc}

The optical depth to reionization, $\tau$, is one of the six parameters of the concordance $\Lambda$CDM
model of cosmology,
and  is given by the line-of-sight integration of the mean electron density $\overline{n_e}$
\begin{equation}
\tau = \sigma_T \int \overline{n_e}(z) \frac{dl}{dz} \,dz\\,
\end{equation}
where $\sigma_T$ is the Thompson cross-section, and $\frac{dl}{dz}$ is the proper line-of-sight distance per unit redshift.
The mean electron density $\overline{n_e}(z)$ can be explicitly calculated using,
\begin{equation}
    \overline{n_e} = \overline{n_b} \left [\overline{x_\textrm{HII}(1+\delta_b)} + \frac{1}{4} \overline{x_\textrm{HeIII}(1+\delta_b)} {Y_p}^\textrm{BBN} \right]\,,
\end{equation}
where $n_b = n_\textrm{H} + n_\textrm{He}$ is the baryon number density, which in turn is the sum of the hydrogen $n_\textrm{H}$
and helium $n_\textrm{He}$ number densities. The ionization fractions (defined to be between $0$ and $1$) are given by $x_\textrm{HII}$, and $x_\textrm{HeIII}$, referring to singly ionized hydrogen and doubly ionized helium, respectively. The helium fraction $Y_p^\textrm{BBN}$ is defined as $4n_\textrm{He} / n_b$, and $\delta_b$ denotes the baryon overdensity. 
Assuming the uncertainties in the helium reionization to be negligible, as in Ref.~\cite{Liu:2015txa}, and 
that helium is instantaneously reionized at $z$ = 3, the expression for the optical depth 
sourced by free electrons from HI/HeI reionization (neglecting the contribution from HeII reionization),
can be simplified as,
\begin{eqnarray}
\label{eq:tau}
\tau = \frac{3 H_0 \Omega_b \sigma_Tc}{8 \pi G m_p} \left[ 1 + \frac{Y_p^\textrm{BBN}}{4}\left( \frac{m_\textrm{He}}{m_\textrm{H}} - 1\right)\right]^{-1} \nonumber \\ 
\times \int_0^{z_\textrm{CMB}} \frac{dz (1+z)^2}{\sqrt{\Omega_\Lambda + \Omega_m (1+z)^3}}  \overline{x_\textrm{HII} (1+\delta_b)}
\end{eqnarray}
Note that, except for the $\overline{x_\textrm{HII}. (1+\delta_b)}$ term, which is the density-weighted
ionization fraction (and is not equal to $\overline{x_\textrm{HII}}.\overline {(1+\delta_b)}$),
all the other terms are either fundamental constants or cosmological parameters constrained by Planck and other observations. 
These measurements of the cosmological parameters come with their own uncertainties, which introduces errors in the
calculation of $\tau$.

So, for the time being, let us focus on this $\overline{x_\textrm{HII}. (1+\delta_b)}$ term. 
Due to the presence of $(1+\delta_b)$, it is not straightforward to calculate $\overline{x_\textrm{HII} (1+\delta_b)}$. 
However, we can guess its values at certain redshifts. For example, at $z\!<\!5$, where the universe is 
almost completely ionized, we can approximately take $\overline{x_\textrm{HII} (1+\delta_b)}\approx1$, which
makes the integration utterly simple in this $z$ range. Meanwhile, at high redshifts (preceding
reionization) we do not have sufficient free electrons, and so  
$\overline{x_\textrm{HII} (1+\delta_b)}\approx0$. Guided by these arguments,  it is sufficient to
take the upper limit of the integration in Eq.~\eqref{eq:tau} to be $z\!=\!35$. Note that this calculation of $\tau$ is subject to
 uncertainties in the cosmological and astrophysical parameters, where the latter are more dominant. Therefore we need to understand and model the astrophysical processes more
precisely in order to have more reliable predictions of $\tau$. 

\begin{figure} [t]
	\centering
	\includegraphics[scale=0.6]{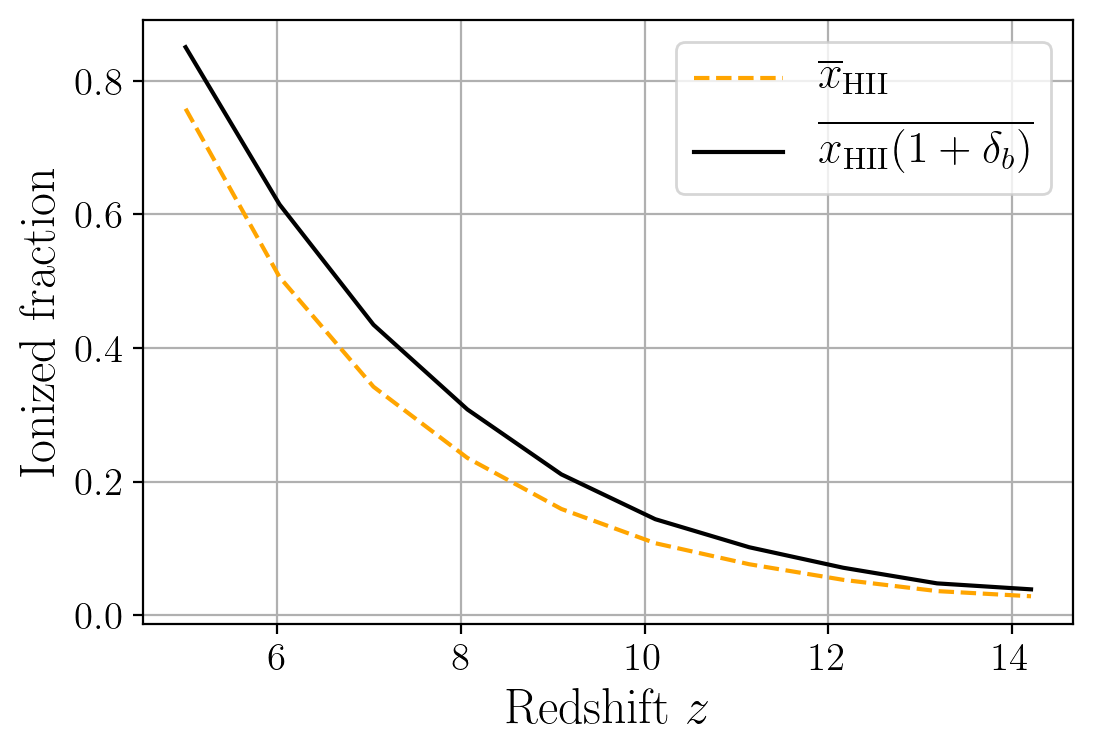}
	\caption{The density-weighted ionized fraction $\overline{x_\textrm{HII}(1+\delta_b)}$,
 together with the mean ionized fraction $\overline{x}_\textrm{HII}$, calculated using the 
 corresponding density and ionized fields generated by the \21cmf simulations considering \cai~for the astrophysics model,
 mentioned in Section~\ref{sec:21cm_obs}. Note here that, 
 $\overline{x_\textrm{HII}(1+\delta_b)}$ and $\overline{x}_\textrm{HII}$ are very 
 different at low redshifts.}
	\label{fig: density-weighted ionized fraction}
\end{figure}

In Figure~\ref{fig: density-weighted ionized fraction}, we show the density-weighted mean $x_\textrm{HII}$
{\it i.e.} $\overline{x_\textrm{HII}. (1+\delta_b)}$ along with $\overline{x_\textrm{HII}}$, calculated
from \21cmf. Both  quantities increase rapidly below $z\sim10$ and tend to approach $\sim1$ at 
low redshifts. Although at high redshifts ($z>14$), $\overline{x_\textrm{HII}. (1+\delta_b)}$
and $\overline{x_\textrm{HII}}$ seem to overlap, we see that they are very different at $z<14$. 
In fact, $\overline{x_\textrm{HII}}$ remains below $\overline{x_\textrm{HII}. (1+\delta_b)}$ in the
range $z<14$. This suggests that if we calculate $\tau$ based on $\overline{x_\textrm{HII}}$ alone, 
we will underpredict the $\tau$ value.

\subsection{Computing $\tau$ from 21-cm observations}

The 21-cm power spectrum is very sensitive to this combination $x_\textrm{HII} (1+\delta_b)$, as 
can be seen from Eq.~\eqref{eq:T21}.
However, $x_\textrm{HII} (1+\delta_b)$ cannot be inferred directly from the 21-cm observations.
The 21-cm power spectrum probes a complicated combination of $x_\textrm{HII} (1+\delta_b)$ and astrophysics 
(through $T_s$) and this introduces degeneracy. Therefore, we not only need a very precise measurement of the
21-cm signal, we also need to have very accurate modeling of the astrophysics in order to use the power spectrum
to probe $\tau$. In order to extract $x_\textrm{HII} (1+\delta_b)$ from the 21-cm power spectrum, 
we use the \21cmf semi-numerical code. 
We use the code to compute the quantity $\overline{x_\textrm{HII} (1+\delta_b)}$ in a light-cone box, 
and integrate it in redshift to get the desired $\tau$.
Note that, the choice of reionization process is important here as it decides the sign of the correlation 
between $x_\textrm{HII}$ and $\delta_b$, and here we assume an inside-out model of reionization.

We now briefly outline the process to determine $\tau$ in an actual scenario where
the 21-cm power spectrum is measured with high 
confidence from any 21-cm observations. Given a measured 21-cm power spectrum, one can use the \21cmf 
code in a Bayesian inference pipeline to simultaneously fit for the astrophysical and cosmological parameters.
Priors on the cosmological parameters can be drawn from CMB or any other observations that measure these 
parameters more precisely. 
Once the parameters are determined with confidence, the \21cmf code can be run to determine 
$\overline{x_\textrm{HII} (1+\delta_b)}$ and calculate $\tau$. Since we are using mock data,
in order to simplify the analysis we use an equivalent Fisher formalism that basically mimics the 
above steps. We discuss the Fisher formalism below.

\subsection{Degeneracy between $\tau$ and model parameters from 21-cm observations}

In this section, we discuss the degeneracy between $\tau$ and other parameters, both astrophysical and cosmological. 
It is important to study the degeneracy in order to understand the improvements on $\Delta \sum m_{\nu}$ due to
21-cm observations. The fact that CMB or any other probe of cosmological parameters will have different 
$\tau - \sum m_{\nu}$ degeneracy in comparison to the 21-cm observations, helps  reduce 
$\Delta \sum m_{\nu}$. We follow Ref. \cite{Liu:2015txa} and explain the process briefly below. 
Unless otherwise indicated, the results presented here are primarily based on \ca. 

We define  $\tau (\textbf{p})_{\rm sim}$ as a function of  $11$ parameters 
$\textbf{p} = [h, \Omega_bh^2, \Omega_ch^2, \ln(10^{10}A_s), n_s, \log L_X, \log f_{\star,10}, 
 \log f_{\rm esc,10}, \allowbreak
\log L_{\rm X,mini}, \log f_{\star,7}, \log f_{\rm esc,7}]$, and
given a set of parameter values,  $\tau (\textbf{p})_{\rm sim}$ yields the value $\tau$ from simulation. 
Note that, the \21cmf simulation allows many more parameters that one can vary in order to calculate $\tau$. 
However, the inclusion of all the parameters makes the analysis lengthy. Furthermore, some parameters do not change
the $\tau$ values significantly even when varied over a significantly large range.  
Based on these considerations, we have chosen the above parameters to study the degeneracies.  
For more details about the astrophysical parameters, the reader is referred to 
Refs.~\cite{Qin:2020xyh,Munoz:2021psm}.

Like in Ref. \cite{Liu:2015txa}, we now seek a linearized relation between $\tau$ and the other parameters for a set of fiducial values. 
Considering a small change $\Delta p$ around the fiducial values, we find the linearised fit to the simulations 
\begin{eqnarray}
\label{eq:tau linear}
\tau_\textrm{sim} \approx &&\, 0.056 + 0.18 \left (\frac{\Delta h}{0.6766}  \right) + 0.019 \left (\frac{\Delta \Omega_bh^2}{0.02242}  \right) \nonumber \\
&& + 0.163 \left (\frac{\Delta \Omega_ch^2}{0.11933}  \right) + 0.12 \left (\frac{\Delta \ln(10^{10}A_s)}{3.047}  \right) \nonumber \\
&& + 0.19 \left (\frac{\Delta n_s}{0.9665}  \right) \nonumber \\
&& + 0.0065 \left (\frac{\Delta \log L_X}{39.0}  \right)  - 0.043 \left (\frac{\Delta \log f_{\star,10}}{-1.45}  \right) \nonumber \\
&& - 0.047 \left (\frac{\Delta \log f_{esc,10}}{-1.42}  \right) + 0.00043 \left (\frac{\Delta \log L_{\rm X,mini}}{39.0}  \right) \nonumber \\
&& - 0.033 \left (\frac{\Delta \log f_{\star,7}}{-3.0}  \right) - 0.016 \left (\frac{\Delta \log f_{esc,7}}{-1.42}\right).
\end{eqnarray}

Note that, $\Delta p$ does not denote the uncertainties in the parameters. Rather, it can be treated as a small 
perturbation around the fiducial values in order to obtain the linearized relation. We have chosen fiducial 
values such that those produce the best-fit $\tau$ value measured by the Planck dataset as discussed in
Section~\ref{sec:cmb_obs}. As mentioned in Ref. \cite{Liu:2015txa}, 
this is needed for the self-consistency of the 21-cm predicted $\tau$ and the CMB optical depth, which we shall discuss next. 
From Eq.~\eqref{eq:tau linear}, it can be seen that the coefficients of the linearized equation for the astrophysical 
parameters are pretty small, and almost an order of magnitude smaller compared 
to the coefficients of some of the cosmological parameters. 
The cosmological parameters, on the other hand, have large degeneracies with
$\tau$. Note that the coefficients of the linearized relation change to some extent with a change in the fiducial
values of the parameters. However, this effect is not huge and the conclusions remain qualitatively the same. 
The above statements also hold true for Scenarios B and C.

\subsection{Elimination of $\tau$ using 21-cm observations}
\label{sec:elimination_of_tau}

We again emphasize the goal of this work, which is to constrain $\tau$ using 
21-cm observations and obtain tighter constraints on  cosmological parameters like $\Delta \sum m_{\nu}$.
In order to do so, we may perform the following likelihood analysis where we incorporate  $\tau$ as estimated 
from the 21-cm observations into the constraints on the parameter set $\textbf{p}$ by performing 
a constrained marginalization over $\tau$ to obtain a likelihood function $\mathcal{L}(\textbf{p})$. The likelihood
function for such an analysis can be written as
\begin{equation}
\label{eq:likelihood with delta}
\mathcal{L} (\mathbf{p}) = \int d\tau \, \mathcal{L}_\textrm{expt} ( \mathbf{p}, \tau) \, \mathcal{P( \tau (\mathbf{p}))}\,,
\end{equation}
where $\mathcal{L}_\textrm{expt}$ is the likelihood function from different experiments, which may include CMB, galaxy surveys, 
21-cm intensity mapping etc. The probability distribution, $\mathcal{P( \tau (\mathbf{p}))}$, acounts for the 
modeling/simulations uncertainties on $\tau$. For simplicity, following Ref.~\cite{Liu:2015txa}, we assume 
$\mathcal{P( \tau (\mathbf{p}))}$ is a Dirac delta function $ \delta^D \!\!\left( \tau_\textrm{sim} (\mathbf{p}) - \tau \right)$. 
This choice ensures that the inferred value for $\tau$ is consistent with 
the one predicted by inputting all the other cosmological parameters into the simulations. 

Note that the usual analysis would start with a prediction of $\tau$ from 21-cm observations, and then use it 
in the CMB analysis to reduce the uncertainties on the cosmological parameters. But since $\tau$ depends on the
cosmological parameters (as can be seen from Eq.~\eqref{eq:tau linear}), the uncertainties can be reduced further
using this fact. To include all the information in our Fisher analysis, 
following Refs.~\cite{Liu:2015txa,Clesse:2012th}, we assume that
$\tau$ values measured from CMB and predicted by 21-cm observations match. 

We can thus simplify the likelihood $\mathcal{L}_\textrm{expt}$
\begin{eqnarray}
\label{modified likelihood}
\mathcal{L}_\textrm{expt} ( \mathbf{p}, \tau) \propto &&\,\, \exp \bigg{[}-\frac{1}{2} \bigg{(}F_{\tau \tau} (\Delta \tau)^2  + \sum_{i \neq \tau} F_{i \tau} \Delta p_i \Delta \tau \nonumber \\
&& + \sum_{j \neq \tau} F_{\tau j} \Delta p_j \Delta \tau  + \sum_{ij \neq \tau}F_{ij} \Delta p_i \Delta p_j \bigg{ )}\bigg{]}, \qquad
\end{eqnarray}
by assuming that it is a multi-variate 
Gaussian distribution with the correlations given by the components of the Fisher matrices calculated for 
the various observations. Here $\Delta p_i$ and $\Delta \tau$ are the deviations of 
$i$th parameter and $\tau$ around their fiducial values (as stated in Eq.~\eqref{eq:tau linear}), 
and $F_{ij}, F_{i \tau}, F_{\tau \tau}$ are elements of the Fisher matrix $\textbf{F}$. For CMB observations
with Planck, \textbf{F} is just the inverse Planck covariance matrix.
We sum this with the calculated Fisher matrix for the 
21-cm power spectrum using  \21cmf, assuming the noise and foregrounds expected for 
observations with  HERA. 
Now we can evaluate the integral in Eq.~\eqref{eq:likelihood with delta}, which is equivalent to substituting $\tau$ with $\tau_\textrm{sim}(\textbf{p})$ in Eq.~\eqref{modified likelihood}. Using the linear approximation to $\tau_\textrm{sim}(\textbf{p})$,  we then get
\begin{equation}
\label{eq:deltaTau}
\Delta \tau = \sum_i a_i \Delta p_i,
\end{equation}
where $a_i$ are the coefficients that match our linearized relation. Substituting this into Eq.~\eqref{modified likelihood} gives us another Gaussian likelihood for $\mathcal{L} (\mathbf{p})$ with modified Fisher matrix
\begin{equation}
\label{eq:modified fisher}
F^\prime_{ij} = F_{ij} + a_i F_{j \tau} + a_j F_{i \tau} + a_i a_j F_{\tau \tau}.
\end{equation}
This Fisher matrix forms the basis of our analysis, where the information on $\tau$ from the 21-cm power spectrum
enters as the coefficients $a_i$. We can now use this modified Fisher matrix to forecast constraints on the parameters.

\section{Results} 
\label{sec:CMBresults}

\subsection{Parameter uncertainties with 21-cm estimated $\tau$}
\label{sec:parameter_uncertainty}

Based on the likelihood function, Eq.~\eqref{modified likelihood}, and the Fisher matrix in Eq.~\eqref{eq:modified fisher}, 
derived in the previous section, we now proceed to make some quantitative predictions. 

As  mentioned previously, the \21cmf code has a number of astrophysical model parameters. 
When we include MCGs in our calculations (Scenario A), the number of astrophysical parameters 
almost doubles in comparison to simulations where only ACGs exist. However, we find that parameters
for MCGs do not change the $\tau$ value considerably even when varied over a large range. Also, the 
fiducial values for these parameters are largely unknown at high redshifts. Based on these, we consider
three cases under Scenario A. \textbf{\cai}: In this case, we vary the ACG parameters $L_X, f_{\star,10}, f_{\rm esc,10}$
along with the cosmological parameters, 
keeping the MCG parameters $L_{\rm X,mini}, \allowbreak f_{\star,7}, f_{\rm esc,7}$ fixed. 
\textbf{\caii}: This is quite the opposite of \cai. We vary the MCG parameters along with the cosmological parameters,
keeping the ACG parameters fixed. 
\textbf{\caiii}: In this case, we vary both the ACG and MCG parameters. 
Since \caiii~has more astrophysical parameters, at the outset, we expect that any forecast 
with \caiii~is likely going to be worse as 
compared to \cai~and~\caii. 

The top (bottom) portion of Table~\ref{tab:astro_forecast1} contains the $1-\sigma$ constraint on the astrophysical parameters
(with fiducial values quoted in the second column) for \cai~(\caii) based on the 
mock measurement of the 21-cm power spectrum using HERA, as well as constraints that arise
from requiring that the parameters self-consistently reproduce $\tau$ in semi-analytic simulations,
which we shall refer to as the `\textbf{self-consistency requirement}' from now on. In parentheses are the $1-\sigma$ constraints deduced from \caiii~for the entire set of astrophysical parameters considered.
Comparing the last two columns, we see that the self-consistency requirement barely improves the
constraints on the astrophysical parameters. This result is expected and congruous with the findings of Ref.~\cite{Liu:2015txa}. 
We also find that the MCG parameters are almost an order of magnitude less constrained compared to ACG 
parameters. This indicates that the variation in the MCG parameters does not change the 21-cm signal 
and $\tau$ values as much as the ACG parameters. Considering \caiii, we found that 
the ACG parameters are slightly less constrained compared to \cai, and the MCG parameter
constraints are very close to what we have
for \caii. This is expected as we have more parameters in \caiii~as compared to the other two scenarios and
this is possibly making the errors larger. 
However, since the constraints in \caiii~are not  much different from 
the other two cases, we shall only consider the results of \cai~and \caii~from here onwards.

\begin{table}
\begin{ruledtabular}
\begin{tabular}{lllll}
Parameter & Fiducial Value & Errors from $P_{21}(k)$ & +21-cm $\tau$ \\ 
\hline
$\log L_{\rm X}$ \dotfill & 39.0  & $\pm0.013$ (0.013) & $\pm0.012$ (0.013) \\
$\log f_{\star,10}$ \dotfill & -1.45  & $\pm0.022$ (0.024) & $\pm0.018$ (0.019) \\ 
$\log f_{\rm esc,10}$ \dotfill & -1.42  & $\pm0.018$ (0.032) & $\pm0.018$  (0.032) \\ \hlinewd{2pt}

$\log L_{\rm X,mini}$ \dotfill & 39.0  & $\pm0.12$  (0.13) & $\pm0.12$  (0.013) \\
$\log f_{\star,7}$ \dotfill & -3.0 & $\pm0.18$  (0.23) & $\pm0.18$  (0.023) \\
$\log f_{\rm esc,7}$ \dotfill & -1.42  & $\pm0.17$  (0.20) & $\pm0.17$  (0.020) \\
\end{tabular}
\end{ruledtabular}
\caption{The fiducial values and $1\;\sigma$ errors for the astrophysical parameters.
The results are produced by simultaneously constraining the astrophysics and cosmology parameters 
while imposing Planck ``TT, TE, EE + LowE + Lensing + BAO" prior (Section~\ref{sec:cmb_obs}) on the latter. 
The ``Errors from $P_{21}(k)$" column shows the $1\;\sigma$ forecast from the 21-cm power spectrum
measured using HERA (Section~\ref{sec:21cm_obs}). The last column contains the errors following the
`self-consistency requirement' (see Section~\ref{sec:elimination_of_tau}), which demands that the 
CMB-measured $\tau$ matches the value of $\tau$ that is predicted from 21-cm observations. 
The first three rows correspond to the astrophysical parameters for \cai, while the last three rows 
correspond to the same for \caii. Inside the parenthesis of all the rows, we show the results for \caiii. 
Note that the self-consistency requirement barely improves the constraints on the astrophysics parameters.}
\label{tab:astro_forecast1}
\end{table}

\begin{table}
\begin{ruledtabular}
\begin{tabular}{llllll}
Parameter & Fiducial Value & Planck & +$P_{21}(k)$ & +21-cm $\tau$ \\
\hline
$H_0$ \dotfill & 67.66 & $\pm0.42$ & $\pm0.15$ & $\pm0.15$ \\
$\Omega_b h^2$ \dotfill & 0.02242 & $\pm0.00014$ & $\pm0.00012$ & $\pm0.00012$ \\
$\Omega_c h^2$ \dotfill & 0.11933 & $\pm0.00093$ & $\pm0.00017$ & $\pm0.00017$ \\
$\ln(10^{10}A_s)$ \dotfill & 3.047 & $\pm0.014$ & $\pm0.012$ & $\pm\boldsymbol{0.0057}$ \\
$n_s$ \dotfill & 0.9665 & $\pm0.0037$ & $\pm0.0028$ & $\pm0.0028$ \\
$\tau$ \dotfill & 0.056 & $\pm0.0072$ & $\pm0.0060$ & $\pm\textbf{\emph{0.0012}}$ \\ 
\end{tabular}
\end{ruledtabular}
\caption{The fiducial values and $1\sigma$ constraints on $\Lambda {\rm CDM}$
cosmological parameters. The third column shows the errors from the Planck 
``TT, TE, EE + LowE + Lensing + BAO" dataset (Section~\ref{sec:cmb_obs}). 
The fourth column shows the $1\sigma$ forecast from the 21-cm power spectrum
measured using HERA gc (considering \cai~for the astrophysics model,
 mentioned in Section~\ref{sec:21cm_obs}). The final column contains the errors following the
`self-consistency requirement' (see Section~\ref{sec:elimination_of_tau}).
The boldfaced entries represent a substantial reduction in error. 
In the last column, $\tau$ is a derived quantity and the corresponding error is written in italics.
Note that the self-consistency requirement significantly improves the constraint on $A_s$.
}
\label{tab:cosmo_forecast1}
\end{table}

\begin{figure} [t]
	\centering
	\includegraphics[scale=0.6]{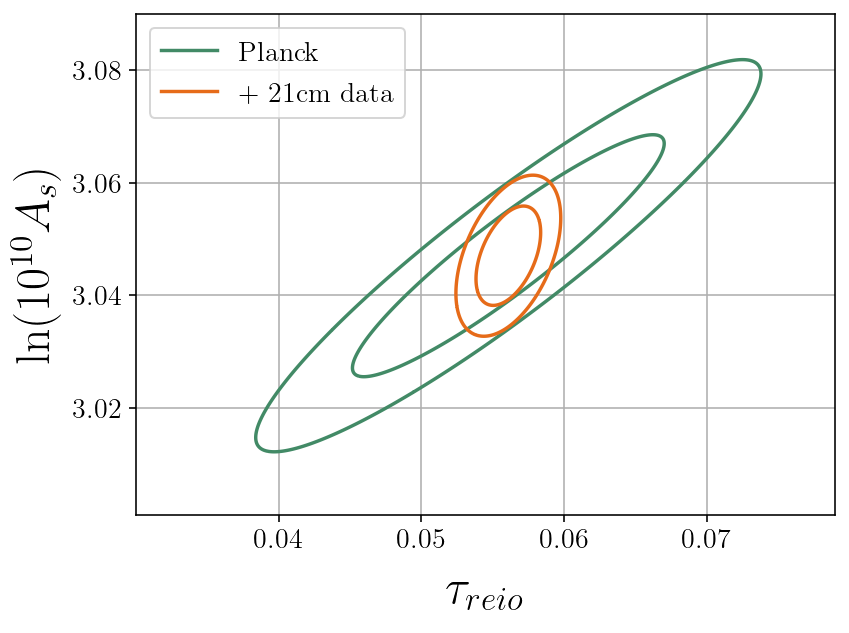}\\
	\caption{The 95\% and 68\% confidence ellipses
 in the $\tau-\ln(10^{10}A_s)$ plane. The green contours denote the constraints from Planck
 ``TT, TE, EE + LowE + Lensing + BAO" data set, while the orange contours are generated after 
 including the 21-cm power spectrum observations from HERA (considering \cai~for the astrophysics model,
 mentioned in Section~\ref{sec:21cm_obs}) and the self-consistency requirement on $\tau$
 (see Section~\ref{sec:elimination_of_tau}). We see that the 21-cm observations and the 
 self-consistency requirement together reduce the $\tau-A_s$ degeneracy and also
 improve the constraints on $A_s$. }
\label{fig: tau-A_s ellipses}
\end{figure}

Considering the same for cosmological parameters in Table~\ref{tab:cosmo_forecast1}, for \cai~alone, we find that there is a significant improvement in the constraints on some of the parameters
when adding the 21-cm power spectrum and most importantly the self-consistency requirement.  
We can see, for example, a noticeable reduction in the error bar for $\ln(10^{10}A_s)$, as 
the 21-cm observations break the CMB degeneracy ($A_s e^{-2 \tau}$) between $A_s$ and $\tau$, 
allowing much better constraints on both parameters. This is evident from Figure~\ref{fig: tau-A_s ellipses},
where the $1$ and $2\,\sigma$ ellipses on the $A_s - \tau$ plane
for Planck shrink and become less tilted when we add information from 21-cm observations. Notice that we had to account for the covariant term between $A_s$ and $\tau$, which was done by considering only the $A_s$ part in Eq.~\ref{eq:deltaTau} and taking $\Delta p$ as the 1$\sigma$ error on $A_s$. This, of course, is the term determining the tilt of the ellipse.
Now considering results for \caii~in Table~\ref{tab:cosmo_forecast1_A2} in Appendix~\ref{sec:tables},
we see that the cosmological parameter constraints are 
almost similar for these two scenarios. Based on this finding, from now onwards, we shall drop \caii~and 
mainly focus on \cai~whenever we discuss results for Scenario A.

As mentioned, in the formalism developed in Section~\ref{sec:elimination_of_tau}, $\tau$ is marginalized 
out of the set of parameters as a self-consistency requirement. Due to this, in Table~\ref{tab:cosmo_forecast1},
$\tau$ is a measured quantity in the third and fourth columns and appears as a derived quantity in the 
last column. Now, to estimate the error on the derived $\tau$, we use the likelihood $\mathcal{L} (\mathbf{p})$
given in Eq.~\eqref{eq:likelihood with delta}. We draw random samples of the parameters $\mathbf{p}$
from $\mathcal{L} (\mathbf{p})$, and feed the values into the linearized relation given by
Eq.~\eqref{eq:tau linear} to calculate $\tau$ for each set of $\mathbf{p}$. The uncertainty on $\tau$
is then estimated from the spread of the $\tau$ values. Note that the random drawing process needs to be
continued until the standard deviation of the $\tau$ distribution converges. We find that for our calculations, 
we needed $\mathcal{O}(1000)$ iterations of random drawings before the standard deviation converged.
Considering the random drawing of 
both astrophysical and cosmological parameters in this manner and focusing on \cai, we find a $1\,\sigma$ error on $\tau$
of $\pm0.0012$. It is interesting to check which set among the astrophysical and cosmological parameters
yield the maximum error on $\tau$. Suppose we first want to check the effect of the cosmological parameters. 
To do so, we marginalize Eq.~\eqref{eq:likelihood with delta} over the astrophysical parameters and
obtain $\mathcal{L} (\mathbf{p})$ which now contains only the cosmological parameters. Given this 
$\mathcal{L} (\mathbf{p})$ in hand, we perform the random drawing of the cosmological parameters. 
We find that the cosmological parameters yield $\pm 0.00052$ error on $\tau$. Repeating the same process
for the astrophysical parameters, where the cosmological parameters are marginalized before obtaining the 
desired $\mathcal{L} (\mathbf{p})$, we find that astrophysical parameters introduce 
$\pm 0.00060$ error on $\tau$. Note that, the error reduces significantly
once we drop any one set of parameters. This suggests that there are degeneracies between astrophysical and
cosmological parameters, which enhances the error when we consider both sets, and the error is reduced
when we fix either set.

\begin{table}
\begin{ruledtabular}
\begin{tabular}{llll}
 & Fiducial & S4$_{\ell > 50}$+Euclid & $+P_{21} (k)$ \\
Parameter & Value & +\emph{Planck} Pol& $+21\,\textrm{cm}$ $\tau$ \\
\hline
$H_0$ \dotfill & $67.66$ & $\pm 0.18$ & $\pm 0.09$ \\
$\Omega_b h^2$ \dotfill & 0.02242 & $\pm 0.000031$ & $\pm 0.000030$ \\
$\Omega_c h^2$ \dotfill  & 0.11933 &$\pm 0.00032$ & $\pm 0.00011$\\
$\ln ( 10^{10} A_s) $ \dotfill  & 3.047 & $\pm0.0078 $ & $\pm0.0009 $\\
$n_s$ \dotfill  & 0.9665 & $\pm 0.0015 $ & $\pm 0.0013$\\
$\tau$ \dotfill  & 0.056  & $\pm 0.0043 $ & $\pm\textbf{\emph{0.00035}}$ \\
$\sum m_\nu$ [meV] \dots & 60 & $\pm 16.2$ & $\pm\textbf{11.8}$\\ 
\end{tabular}
\end{ruledtabular}
\caption{
The fiducial values and $1\sigma$ constraints for the usual $\Lambda {\rm CDM}$
cosmological parameters plus the sum of neutrino masses $\sum m_\nu$ as an additional parameter,
as discussed in Section~\ref{sec:neutrino_mass_measurement}. 
The third column shows the $1\sigma$ errors combining the Planck polarization measurement at low $\ell$,
forecast from CMB S4 observations at $\ell \geq 50$ and forecast from galaxy power 
spectrum measured by Euclid survey. The last column shows the $1\sigma$ errors when we add 
forecast for the 21-cm power spectrum measured using HERA (considering \cai~for the astrophysics model,
 mentioned in Section~\ref{sec:21cm_obs}) and for the self-consistency requirement, together
with the data used to produce results in column three.  
The boldfaced entries represent a substantial reduction in error. 
We see that the 21-cm observations together with the self-consistency requirement improve the 
constraints on $\sum m_\nu$.
}
\label{tab:S4CosmoParams}
\end{table}

\subsection{Improvement on the sum of neutrino masses}
\label{sec:neutrino_mass_measurement}

In this section, we show how  information on $\tau$ from 21-cm observations can  reduce the 
uncertainties on the sum of neutrino masses $\sum m_{\nu}$. 
Cosmic neutrinos affect both the cosmic expansion and the evolution of density perturbations. 
In the early Universe, neutrinos are relativistic and behave as radiation. As the Universe cools,
they gradually become non-relativistic and behave like matter. If we consider that the total matter density today
has a contribution from non-relativistic neutrinos, then in the past when neutrinos were relativistic, this contribution
was missing. This leads to a later matter-radiation equality when neutrinos were still relativistic. 
Further, the large thermal velocities of neutrinos allow them to stream out of the dark matter potential wells. 
As a result, they do not contribute to matter clustering, and the growth of structure is suppressed on 
scales smaller than their free-streaming scale. This suppression can be mimicked by a lower 
value of $A_s$. Hence, the degeneracy between $\tau$ and $A_s$ leads to a $\tau-\Sigma m_{\nu}$ degeneracy.

\begin{figure} [!]
	\centering
    \includegraphics[scale=0.6]{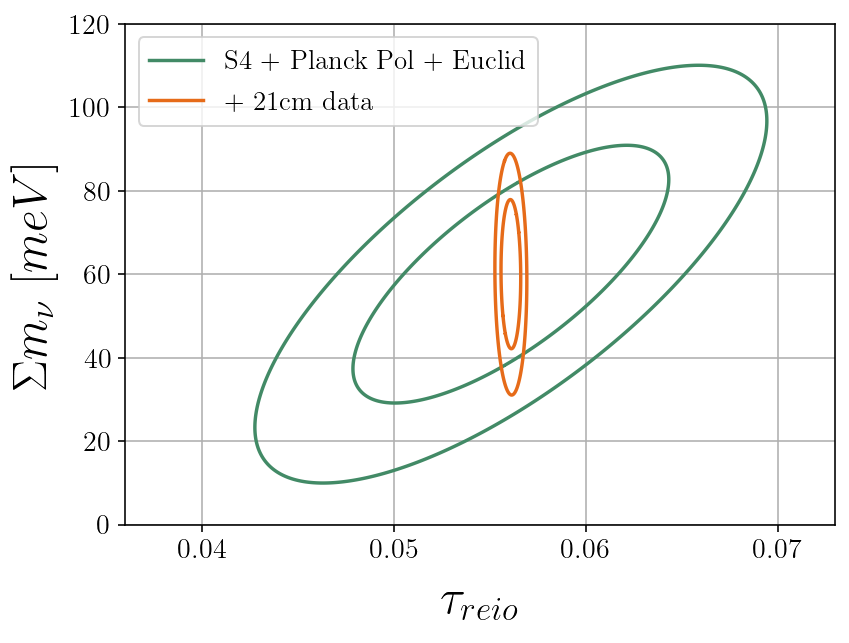}\\
	\caption{The 95\% and 68\% confidence ellipses in the $\tau-\Sigma m_{\nu}$ plane. 
 The green contours show the results when we combine the Planck polarization measurement at low $\ell$,
 the CMB-S4 forecast at $\ell \geq 50$ and the Euclid galaxy power 
spectrum forecast. The orange contours show results when, in addition to the 
data used for the green contours, we combine the 21-cm power spectrum forecast for HERA (considering \cai~for the astrophysics model, mentioned in Section~\ref{sec:21cm_obs}) and the self-consistency requirement on $\tau$
(see Section~\ref{sec:elimination_of_tau}). We see that the 21-cm observations and the 
 self-consistency requirement together diminish the $\tau-\Sigma m_{\nu}$ degeneracy and also
 improve the constraints on $\Sigma m_{\nu}$.
 }
\label{fig: tau and neutrino mass ellipses}
\end{figure}

In order to show how 21-cm observations break $\tau-\Sigma m_{\nu}$ degeneracy, we repeat the same analysis
used to produce the results of Section~\ref{sec:parameter_uncertainty}, but this time including $\Sigma m_{\nu}$ as an
additional cosmological parameter. The results are presented in Figure~\ref{fig: tau and neutrino mass ellipses} and Table~\ref{tab:S4CosmoParams}. For the future CMB-S4
predictions (details provided in Appendix~\ref{sec:cmbs4}), we use the multipoles $\ell=50$ and above. 
Measurement at these multipoles will be important to break the degeneracy. 

Forecasts for
the Euclid galaxy survey are generated following Ref.~\cite{Yankelevich:2018uaz}. 
Here we use the constraints coming only from the
galaxy power spectrum. Note that, it is possible to also include the constraints coming from the galaxy
bispectrum which will likely help to reduce the parameter errors
\cite{Bharadwaj:2020wkc,Mazumdar:2020bkm,Shaw:2021pgy,Mazumdar:2022ynd,Yankelevich:2018uaz,Gualdi:2020ymf}. 
However, for the current analysis, the forecasts 
from the galaxy power spectrum are sufficient. Planck measurement estimated
$\Delta \Sigma m_{\nu}$ to be $ \pm 38\,{\rm meV}$. With  CMB-S4 observations at low $\ell$, combined
with future results from Euclid, it is possible to reduce the $1\,\sigma$
uncertainty down to $\pm 16.2\,{\rm meV}$. Now considering \cai, when we combine the results from  future HERA
observations with the self-consistency requirement, we find that the uncertainty on $\Sigma m_{\nu}$ 
shrinks further down to $\pm 11.8\,{\rm meV}$. This uncertainty is even smaller, $\pm 6.8\,{\rm meV}$, in the case 
of \caii, as shown in Table~\ref{tab:S4CosmoParams_A2} in Appendix~\ref{sec:tables}. The $\tau-\Sigma m_{\nu}$ degeneracy breaking due to 21-cm observations 
can also be seen in Figure~\ref{fig: tau and neutrino mass ellipses} where the tilt of the ellipses 
becomes vanishingly small as we add information from reionization via the 21-cm observation. 
The ellipses also shrink, indicating that the degeneracy 
breaking further improves the constraints on the parameters. In summary, the analysis in this section indicates that 
it is possible to weigh the neutrinos with $\gtrsim 5\;\sigma$
accuracy if we combine 21-cm and CMB data.

\section{Comparing With Other Scenarios}
\label{sec:scenarios}

In this section, we compare the results for \ca~with \cb~and \cc. Note again, that both \cb~and 
\cc~consider only ACGs in the calculations. The only difference is in the parametrization. While 
\cb~uses the latest parametrization, \cc~uses the old parametrization of \21cmf~\cite{Liu:2015txa}. 

Considering Table~\ref{table:Astro_params2}, we find that the ACG parameters for
both \ca~and \cb~are similarly constrained from the 21-cm observations and the self-consistency
requirement improves the constraints only slightly. In the case of \cc, the astrophysical
parameters are pretty well constrained and the self-consistency requirement makes a small improvement.

Comparing the results with Ref.~\cite{Liu:2015txa}, we find that in our case $\zeta$ and $T_{\rm vir}$ 
are slightly better constrained, and $R_{\rm mfp}$ is slightly less constrained. Note that,
we have used $30$ redshift bins in the range $z=5$ to $27$. On the other hand, Ref.~\cite{Liu:2015txa}
 used a limited redshift range $z\sim6$ to $\sim10$, albeit with many more
redshift bins within this range. As we have stated earlier, observations at different redshifts
help to break the degeneracy between the different astrophysics parameters and ultimately reduce the
uncertainty on the parameters. This is possibly happening with $\zeta$ and $T_{\rm vir}$ in our analysis.
Now, the parameter $R_{\rm mfp}$ is only important during the reionization. The analysis
in Ref.~\cite{Liu:2015txa} mainly focuses on this $z$ range and has many more $z$ bins than ours on this particular range.
Therefore, they have more information on $R_{\rm mfp}$ which helps  reduce the uncertainty 
in $R_{\rm mfp}$ in their analysis. 

\begin{table}
\begin{ruledtabular}
\begin{tabular}{lllll}
Parameter & Fiducial Value & Errors from $P_{21}(k)$ & +21-cm $\tau$ \\ 
\hline
$L_{\rm X}$ \dotfill & 39.0  & $\pm0.013$ & $\pm0.012$ \\
$f_{\star,10}$ \dotfill & -1.45  & $\pm0.025$ & $\pm0.017$ \\ 
$f_{\rm esc,10}$ \dotfill & -1.42  & $\pm0.013$ & $\pm0.010$ \\ \hlinewd{2pt}
$\zeta$ \dotfill & 30.0 & $\pm 1.02$ & $\pm 1.02$ \\
$T_{vir}$ [K] \dotfill & $8.5\times 10^4$ & $\pm 2.21\times10^3$ & $\pm 2.13\times10^3$ \\
$R_{mfp}$ [Mpc] \dotfill & 35 & $\pm 6.5$ & $\pm 6.4$ \\
\end{tabular}
\end{ruledtabular}
\caption{Same as Table~\ref{tab:astro_forecast1}, but for \cb~(top rows) and \cc~(bottom rows).
}
\label{table:Astro_params2}
\end{table}

In Table~\ref{tab:scenarios1}, we compare the constraints on the cosmological parameters
for the three different scenarios of the 21-cm signal modeling.
We find that in all the scenarios, the cosmological parameters have similar errors
when we add the 21-cm power spectrum observations. For all the scenarios, 
the self-consistency requirement improves the constraints on $A_s$ significantly, 
and the constraints on other parameters are improved marginally. Despite this, for
each parameter, the constraints look similar in all the scenarios. This is true even 
for the derived parameter $\tau$, and we find the minimum $\tau$ error for \cb~and maximum for \cai. Overall, the results in Table~\ref{tab:scenarios1}
suggest that the constraints on the cosmological parameters weakly depend on the detailed modeling of  
the astrophysics. Note that the constraints can change for different fiducial values of the parameters,
although the qualitative results will remain the same. 

\begin{figure}[!]
     \centering
     \begin{subfigure}{}
         \centering
         \includegraphics[scale=0.55]{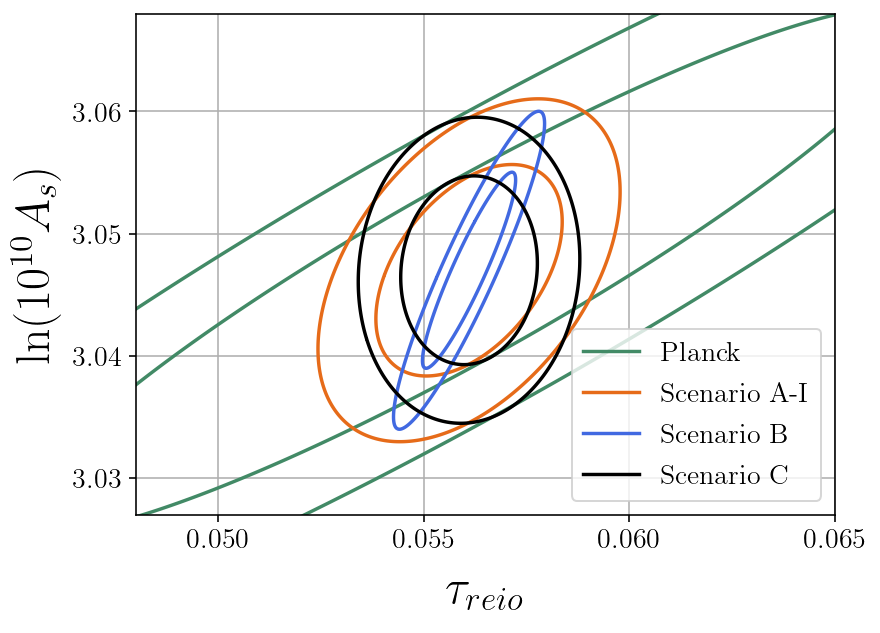}
         \label{fig:y equals x}
     \end{subfigure}
     \begin{subfigure}{}
         \centering
         \includegraphics[scale=0.55]{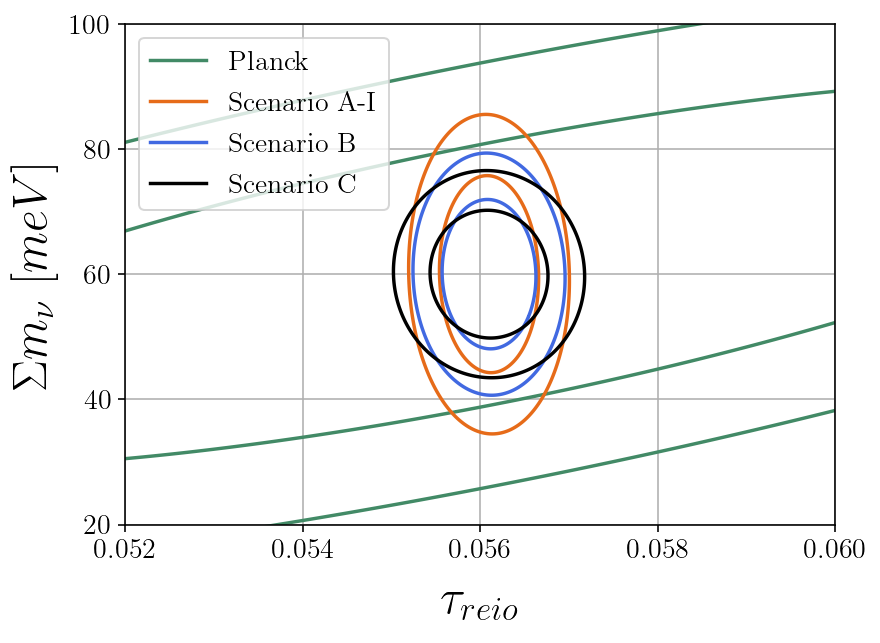}
         \label{fig:three sin x}
     \end{subfigure}
        \caption{Top panel is same as Figure~\ref{fig: tau-A_s ellipses}  and 
        bottom panel is same as Figure~\ref{fig: tau and neutrino mass ellipses} (bottom). 
        In addition to \cai, 
        these figures show results for \cb~and \cc. We see that the $\tau-A_s$ degeneracy is minimum 
        for \cc.
        We see that the $\tau-\Sigma m_{\nu}$ degeneracy is
        vanishingly small for all  scenarios.}
        \label{fig:ellipses 3 scenarios}
\end{figure}

In Figure~\ref{fig:ellipses 3 scenarios} (top), we compare the $\tau-A_s$ degeneracy breaking for the
three scenarios mentioned above. Comparing the tilts of the ellipses in  the $\tau-A_s$ plane, we see that 
the tilt is maximum for \cai, followed by \cb, and we find almost no tilt for \cc. The errors on $\tau$
and $A_s$ are also largest for \cai. This indicates that, in certain astrophysical scenarios,
the $\tau-A_s$ degeneracy cannot be removed fully even with  information on $\tau$ supplied
from  the 21-cm observations, and the degree of this remaining degeneracy is model dependent.

In Table~\ref{tab:scenarios2}, we compare our results for $\Delta \sum m_\nu$ under  
the three astrophysical scenarios. In column two of this table, we show predictions for $\Delta \sum m_\nu$
when the Planck polarization (low $\ell$) data is analyzed together with CMB-S4 and Euclid data. 
Note that CMB-S4 is predicted to measure the lower power spectrum multipoles very accurately,
which is crucial for the measurement of the reionization bump and this provides an independent estimate
of $\tau$. We find that $\Delta \sum m_\nu$ is smallest for \cc, 
followed by \cb, and finally we get maximum $\Delta \sum m_\nu$ ($\pm 11.8\,[{\rm meV}]$)
for \cai~which is slightly better than the result obtained by combining the estimates 
of the Planck polarization map along with the CMB-S4 and Euclid predictions ($\pm 16.2\,[{\rm meV}]$). Although the
derived parameter $\tau$ has maximum error for \cc, the constraints on $\tau$ for other scenarios
are not far from this value. These results indicate that the 
accuracy of the neutrino mass measurement depends very much on the modeling of astrophysics.

In Figure~\ref{fig:ellipses 3 scenarios} (bottom), we show the $\tau-\Sigma m_{\nu}$ degeneracies
for the three astrophysical scenarios. Comparing the tilt of the ellipses, we can clearly see that
the degeneracy is vanishingly small for all three scenarios considered. This suggests that the 
21-cm observations can successfully break the degeneracy between $\tau$ and $\Sigma m_{\nu}$, and this
result is largely independent of the modeling of astrophysics.

\section{Summary and Conclusions}
\label{sec:summary}

Neutrino oscillation experiments have established that neutrinos have non-zero mass, and there is a 
mass hierarchy for the three flavors of neutrinos. Massive neutrinos change
the kinematics of our Universe’s expansion. They also dampen structure growth on scales
below their free-streaming length,  leading to a deficit in power on small scales. 
This effect is more pronounced if the sum $\Sigma m_{\nu}$ of their masses  is large. 
Neutrinos signatures can be observed in the CMB and galaxy power spectra, and 
$\Sigma m_{\nu}$ can be constrained using the CMB and large-scale structure observations. 
This, however, is not straightforward. 

The presence of free electrons in the IGM during and after the reionization era limits  
the cosmological measurement of $\Sigma m_{\nu}$. Specifically, the optical depth parameter $\tau$, which
is a measure of the column density of free electrons in the IGM and  one of the $\Lambda$CDM 
model parameters, is degenerate with $\Sigma m_{\nu}$
and the scalar amplitude $A_s$. This degeneracy hinders the precise measurement of
$\Sigma m_{\nu}$, and also of $A_s$. We, therefore, need independent and precise 
estimates of $\tau$ from other observations, which  can then be used to reduce the uncertainty on 
$\Sigma m_{\nu}$. The other option is to measure $\Sigma m_{\nu}$ directly from 
some observations. This paper focused on the former approach. 

The 21-cm signal from reionization is a plausible probe of the IGM electron density, and thus provides
an independent measurement of $\tau$. Using it to mitigate the 
$\tau$-$\Sigma m_{\nu}$ degeneracy was first proposed and studied in
Ref.~\cite{Liu:2015txa}, which showed that combining  21-cm and CMB data can significantly reduce the uncertainty
in $\Sigma m_{\nu}$. In this paper, we revisited this analysis, relaxing sum of the assumptions (such as $\Ts>>\Tcmb$) 
and using a modified version of the \21cmf code that is interfaced directly with the CMB code {\tt CLASS} so that 
the degeneracies between astrophysical and cosmological parameters could be consistently accounted for. 
We have further considered 
additional effects like the \lya heating of IGM, Population III stars, the relative velocity 
between dark matter and baryon fluid, inhomogeneous Lyman-Werner (LW) radiation feedback in our 
analysis. 

We considered  21-cm observations with the HERA radio telescope spanning  the range $5 \le z \le 27$, in 
a scenario with moderate foreground contamination (see Ref.~\cite{Pober:2012zz}). Although the moderate 
foregrounds plague the 21-cm power spectrum measurement at $z>20$, we still have significant
sensitivity near the reionization redshifts. The power spectrum is then used to constrain the 
astrophysical model parameters. The current generation of 21-cm observations, like HERA, 
does not have much sensitivity to constrain the cosmological parameters. Thus, initially, it can be assumed that
the cosmological parameters are known from CMB and other observations, and the astrophysical parameters 
will only be determined from the 21-cm power spectrum. Having the astrophysical parameters in hand, 
we can use them in a simulation, like \21cmf, to generate the density and ionization fields.
$\tau$ can be estimated from the density-weighted ionization fraction. Finally, this $\tau$ value 
can be used in the CMB analysis to break the $\tau - \Sigma m_{\nu}$ degeneracy. 

Realistically the CMB, 21-cm and other data should be analysed jointly in Bayesian analysis to 
predict the astrophysical and cosmological parameters. In this way, the information from the 21-cm observations
propagates self-consistently into predicting the $\tau$ values in the analysis. However, here
we do not perform a joint analysis. Therefore, to achieve this self-consistency, we  assume that
the CMB predicted $\tau$ matches with the $\tau$ estimated from the 21-cm power spectrum. Based on this
self-consistency requirement, following Ref.~\cite{Liu:2015txa}, we devise a Fisher analysis technique and 
summarize our main findings below. 

We find that the astrophysical parameters can be well constrained with  21-cm observations. Although, 
parameters associated with the MCGs are less constrained compared to the parameters related to 
ACGs. MCGs (or Population III stars) do not play a dominant role in the reionization process, so the
parameters related to MCGs do not change the $\tau$ values noticeably even when changed over a large range. 
Considering the forecast for the cosmological parameters, we find that the 21-cm derived $\tau$ information
reduces the uncertainties in all the parameters. This is most significant for  $A_s$, where this mitigates 
the degeneracy $A_s e^{-2\tau}$ inherent to the CMB. However, this degeneracy breaking 
depends on the astrophysics models and we found that in certain astrophysical scenarios, this degeneracy 
is not completely alleviated. 

21-cm observations are key for a precise cosmological measurement of the sum of neutrino masses. Future CMB observations promise to measure the reionization bump at low $\ell$ multipoles of the polarization
power spectrum, which can provide an independent measurement of $\tau$ and break the 
$\tau-\Sigma m_{\nu}$ degeneracy. Meanwhile, future galaxy surveys will provide independent measurements 
of $\Sigma m_{\nu}$. Combining estimates for CMB-S4 and Euclid, along with Planck polarization data, 
we find that $\Sigma m_{\nu}$ can only be measured with a $\pm16.2\,[{\rm meV}]$ error, 
for a fiducial $\Sigma m_{\nu}$ of $60\,[{\rm meV}]$ (the minimum value predicted by terrestrial experiments). 
This uncertainty can be further reduced to 
$\pm 11.8\,[{\rm meV}]$ or lower, using the 21-cm derived $\tau$ information from HERA observations. All this translates into a $\gtrsim 5 \sigma$
detection of $\Sigma m_{\nu}$ for the fiducial $\Sigma m_{\nu}$ value and astrophysical models considered.

Our result is marginally more optimistic than that obtained in Ref.~\cite{Liu:2015txa}.
Considering the bounds from the neutrino oscillation experiments \cite{ParticleDataGroup:2014cgo},
the minimum value of 
$\Sigma m_{\nu}$ is $\sim 60$ meV for the normal hierarchy and $\sim 100$ meV for inverted hierarchy.
Therefore, our results demonstrate that incorporating 21-cm 
data will enable 
a robust determination of the neutrino mass hierarchy.

There are a number of probes, like the \lya forest \cite{Weinberg:2003eg,Wyithe:2004jw,Fan:2005es}, 
\lya emitting galaxies \cite{Malhotra:2004ef,Jensen:2012uk,Dijkstra:2014xta,Dijkstra:2015jdy},
the kSZ effect \cite{Ma:2001xr,McQuinn:2005ce,Park:2013mv,Gorce:2020pcy,Gorce:2022cvb},
etc; that can directly measure $\tau$. 
Experiments like JWST also probe astrophysics during the pre-reionization era, which
again provides ample information on $\tau$ \cite{Robertson:2021ljt}.
The galaxy surveys \cite{Villaescusa-Navarro:2017mfx,Palanque-Delabrouille:2019iyz, Ivanov:2019hqk, Chudaykin:2019ock},
measurement of the expansion rate using distance ladders \cite{Wyman:2013lza,Riess:2021jrx},
\lya forest surveys \cite{Palanque-Delabrouille:2014jca,Baur:2017fxd}, 
line-intensity mapping \cite{MoradinezhadDizgah:2021upg,Bernal:2022jap,Libanore:2022ntl},
the post-reionization 21-cm signal \cite{Villaescusa-Navarro:2015cca,Oyama:2015gma,Pal:2016icc,Sarkar:2016lvb,
Sarkar:2018gcb,Sarkar:2019nak,Sarkar:2019ojl}, etc., provide 
direct estimates of $\Sigma m_{\nu}$. The velocity acoustic oscillations 
\cite{Munoz:2019rhi,Munoz:2019fkt,Sarkar:2022mdz}
(originating from the fluctuations in the relative velocity field between dark matter and baryons) 
are expected to be observed in the 21-cm power spectrum during the cosmic dawn epoch. 
The phase shift in the velocity acoustic oscillations caused
by the supersonic propagation of neutrinos (similar to what is recently constrained from
Baryon Acoustic Oscillations \cite{Baumann:2019keh}) can also provide direct measurement of $\Sigma m_{\nu}$.
All these independent observations can be combined, along with the
CMB and 21-cm data, to pinpoint $\Sigma m_{\nu}$. We shall explore some of these possibilities in future works. 

We finally reiterate the essence of this paper, which is that the high-precision measurement 
of the 21-cm signal from experiments like
HERA and SKA will not only lead to a significant improvement in our comprehension of  
high redshift astrophysics, but also supply invaluable information about cosmology and fundamental physics, such as the sum of neutrino masses.

\section*{acknowledgements}
We thank Tal Adi and Jordan Flitter for useful discussions. and assistance with the Fisher forecasts.
We especially thank Adrian Liu for illuminating discussions, clarifying some of the details in the work 
upon which this paper is based. GS is supported by an M.Sc.\ fellowship 
for female students in hi-tech fields, awarded by the Israeli Council for Higher Eduction. 
EDK acknowledges support from an Azrieli Faculty Fellowship.

\appendix

\section{Fisher analysis formalism for CMB-S4}
\label{sec:cmbs4}
Here we outline the formalism used to calculate the Fisher matrix for CMB-S4 based on Ref.~\cite{Munoz:2016owz}.
The CMB power spectra can be written as
\begin{equation}
    C_l^{XY} = (4\pi)^2 \int k^2 T_l^X(k) T_l^Y(k) P(k) \,dk\
\end{equation}
where the indices $X, Y = {T,E}$ stand for temperature and E-mode polarization respectively, and $T_l^X$ are their transfer functions.
We note we haven't considered the lensing potential $C_l^{dd}$ in our analysis for simplicity.

For a set of parameters $\theta_i$ for which we want to forecast the errors, we define the Fisher matrix as
\begin{equation}
    F_{ij} = \Sigma_l \frac{2l+1}{2} f_{sky} Tr \left[C_l^{-1} \frac{\partial C_l}{\partial \theta_i} C_l^{-1} \frac{\partial C_l}{\partial \theta_j} \right]
\end{equation} \\
where $f_{sky}$ is the covered fraction of sky and the matrix 
$C_l$ is defined as 
\begin{equation}
    \begin{pmatrix}
    \Tilde{C}_l^{TT} & C_l^{TE} \\
    C_l^{TE} & \Tilde{C}_l^{EE} 
    \end{pmatrix} \,.
\end{equation}
We further define 
\begin{eqnarray}
\Tilde{C}_l^{TT} = C_l^{TT} + N_l^{TT}, \\ \nonumber
\Tilde{C}_l^{EE} = C_l^{EE} + N_l^{EE}
\end{eqnarray}
where $N_l^{XX}$ are the noise power spectra, given by
\begin{eqnarray}
N_l^{TT} &=& \Delta_T^2 \exp^{l(l+1)\sigma_b^2}, \\ \nonumber
N_l^{EE} &=& 2 \times N_l^{TT} \,,
\end{eqnarray}
where $\Delta _T$ is the temperature sensitivity and $\sigma_b = \theta_{\rm FWHM}/\sqrt{8\log(2)}$, with the full-width-half-maximum $\theta_{\rm FWHM}^2$ given in radians.

In our analysis we choose the CMB parameters $\theta_i \in \{H_0, \Omega_bh^2, \Omega_ch^2, \ln(10^{10}A_s), n_s, \tau\}$.

\section{Additional results}
\label{sec:tables}

In this section, we present some additional tables that allow a comparison of the results between the different 
astrophysical scenarios. 
Tables~\ref{tab:cosmo_forecast1_A2} and \ref{tab:S4CosmoParams_A2} are same as 
Tables~\ref{tab:cosmo_forecast1} and \ref{tab:S4CosmoParams} respectively, but for \caii. Moreover,
Tables~\ref{tab:scenarios1} and \ref{tab:scenarios2} exhibit a comparison between \cai, \cb~and \cc.

\begin{table}[h!]
\begin{ruledtabular}
\begin{tabular}{llllll}
Parameter & Fiducial Value & Planck & +$P_{21}(k)$ & +21-cm $\tau$ \\
\hline
$H_0$ \dotfill & 67.66 & $\pm0.42$ & $\pm0.24$ & $\pm0.24$ \\
$\Omega_b h^2$ \dotfill & 0.02242 & $\pm0.00014$ & $\pm0.00012$ & $\pm0.00012$ \\
$\Omega_c h^2$ \dotfill & 0.11933 & $\pm0.00093$ & $\pm0.00036$ & $\pm0.00036$ \\
$\ln(10^{10}A_s)$ \dotfill & 3.047 & $\pm0.014$ & $\pm0.011$ & $\pm\boldsymbol{0.0065}$ \\
$n_s$ \dotfill & 0.9665 & $\pm0.0037$ & $\pm0.0030$ & $\pm0.0030$ \\
$\tau$ \dotfill & 0.056 & $\pm0.0072$ & $\pm0.0055$ & $\pm\textbf{\emph{0.0018}}$ \\
\end{tabular}
\end{ruledtabular}
\caption{ 
Same quantities as in Table~\ref{tab:cosmo_forecast1}, here considering \caii~for the astrophysics model (Section~\ref{sec:21cm_obs}).
}
\label{tab:cosmo_forecast1_A2}
\end{table}

\begin{table}[h!]
\begin{ruledtabular}
\begin{tabular}{llll}
 & Fiducial & S4$_{\ell > 50}$+Euclid & $+P_{21} (k)$ \\
Parameter & Value & +\emph{Planck} Pol& $+21\,\textrm{cm}$ $\tau$ \\
\hline
$H_0$ \dotfill & $67.66$ & $\pm 0.18$ & $\pm 0.09$ \\
$\Omega_b h^2$ \dotfill & 0.02242 & $\pm 0.000031$ & $\pm 0.000030$ \\
$\Omega_c h^2$ \dotfill  & 0.11933 &$\pm 0.00032$ & $\pm 0.00014$\\
$\ln ( 10^{10} A_s) $ \dotfill  & 3.047 & $\pm0.0078 $ & $\pm0.0010 $\\
$n_s$ \dotfill  & 0.9665 & $\pm 0.0015 $ & $\pm 0.0013$\\
$\tau$ \dotfill  & 0.056  & $\pm 0.0043 $ & $\pm\textbf{\emph{0.00041}}$ \\
$\sum m_\nu$ [meV] \dots & 60 & $\pm 16.2$ & $\pm\textbf{6.8}$\\
\end{tabular}
\end{ruledtabular}
\caption{
Same quantities as in Table~\ref{tab:S4CosmoParams}, here  considering \caii~for the astrophysics model (Section~\ref{sec:21cm_obs}).
}
\label{tab:S4CosmoParams_A2}
\end{table}


\begin{table*}[]
\begin{ruledtabular}
\begin{tabular}{lllcllcllcll}
 & \multicolumn{3}{c}{ } & \multicolumn{2}{c}{\textbf{\cai}} && \multicolumn{2}{c}{\textbf{\cb}}  && \multicolumn{2}{c}{\textbf{\cc}} \\
 & Fiducial  & $Planck$ && $+P_{21} (k)$ &$+21\,\textrm{cm}$ $\tau$&& $+P_{21} (k)$ &$+21\,\textrm{cm}$ $\tau$&&$+P_{21} (k)$ &$+21\,\textrm{cm}$ $\tau$\\
\hline
\multicolumn{6}{l}{Parameters} \\
$H_0$\dotfill  & $67.66 $&$\pm 0.42$ && $\pm 0.15$ &  $\pm 0.15$ && $\pm 0.13 $&$\pm 0.13$ && $\pm 0.16$ & $\pm 0.16$ \\
$\Omega_b h^2$ \dotfill & $0.02242$&$ \pm 0.00014$ && $\pm 0.00012$ &  $\pm 0.00012$ && $\pm0.0.00012 $&$\pm 0.00012$ && $\pm 0.00012$ &  $\pm 0.00012$ \\
$\Omega_c h^2$ \dotfill & $0.11933$&$ \pm 0.00093$  && $\pm 0.00017$ &  $\pm 0.00017$ && $\pm0.00012$&$ \pm 0.00012$ && $\pm 0.00025$ &  $\pm 0.00025$ \\
$\ln ( 10^{10} A_s) $ \dotfill & $3.047 $&$\pm 0.014$ && $\pm 0.012$ &  $\mathbf{\pm 0.0057}$ & &$\pm 0.012$&$ \mathbf{\pm 0.0053}$ && $\pm 0.010$ & $\mathbf{\pm 0.0052}$  \\
$ n_s $\dotfill  & $ 0.9655$&$ \pm 0.0037$ && $\pm 0.0028$ &  $\pm 0.0028$ && $\pm 0.0024$&$ \pm 0.0024$ && $\pm 0.0030$ &  $\pm 0.0030$ \\
$ \tau $ \dotfill & $0.056 $&$\pm 0.0072$ && $\pm 0.0060$ &  $\pm \emph{0.0012}$  && $\pm0.0062 $&$\mathbf{\pm \emph{0.0007}}$ && $\pm 0.0052$ &  $\mathbf{\pm \emph{0.001}}$ \\
\end{tabular}
\end{ruledtabular}
\caption{Same quantities as in Table~\ref{tab:cosmo_forecast1}, including a comparison between \cai, \cb~and \cc.
}
\label{tab:scenarios1}
\end{table*}

\begin{table*}[]
\begin{ruledtabular}
\begin{tabular}{lllclclcl}
 & \multicolumn{3}{c}{ } & \multicolumn{1}{c}{\textbf{\cai}} && \multicolumn{1}{c}{\textbf{\cb}}  && \multicolumn{1}{c}{\textbf{\cc}} \\
& Fiducial & S4$_{\ell > 50}$+Euclid &&  &&  && 
\\ & Value & +\emph{Planck} Pol && +21-cm data && +21-cm data && +21-cm data \\
\hline
\multicolumn{7}{l}{Parameters} \\
$H_0$\dotfill  & $67.66 $&$\pm 0.18$ && $\pm 0.09$ && $\pm 0.07 $ && $\pm 0.05$ \\
$\Omega_b h^2$ \dotfill & $0.02242$&$ \pm 0.000031$ && $\pm 0.000030$ && $\pm 0.000030$ && $\pm 0.000030$ \\
$\Omega_c h^2$ \dotfill & $0.11933$&$ \pm 0.00032$  && $\pm 0.00011$ && $ \pm 0.00007$ && $\pm 0.00015$ \\
$\ln ( 10^{10} A_s) $ \dotfill & $3.047 $&$\pm 0.0078$ && $\pm 0.0009$ &&$\pm 0.0009$ && $\pm 0.0009$  \\
$ n_s $\dotfill  & $ 0.9655$&$ \pm 0.0015$ && $\pm 0.0013$ && $\pm 0.0012$ && $\pm 0.0011$ \\
$ \tau $ \dotfill & $0.056 $&$\pm 0.0043$ && $\pm\emph{0.00035}$ && $\pm\emph{0.00036}$  &&  $\pm\emph{0.00044}$ \\
$ \sum m_\nu [meV] $ \dotfill & $60$ &$\pm 16.2$ && $\mathbf{\pm 11.8}$ && $\mathbf{\pm 7.9}$ &&  $\mathbf{\pm 6.8}$ \\
\end{tabular}
\end{ruledtabular}
\caption{Same quantities as in Table~\ref{tab:S4CosmoParams}, including a comparison between \cai, \cb~and \cc.}
\label{tab:scenarios2}
\end{table*}


\end{document}